\documentstyle[pra,aps,epsf]{revtex}
\begin{document}
\draft
\author{V.B. Svetovoy\thanks{%
E-mail: svetovoy@nordnet.ru} \ and M.V. Lokhanin}
\address{Department of Physics, Yaroslavl State University and \\
Institute of Microelectronics RAS \\ Sovetskaya 14, Yaroslavl 150000, Russia}
\title{Temperature correction to the Casimir force in cryogenic range \\
and anomalous skin effect }
\date{\today}
\maketitle

\begin{abstract}
Temperature correction to the Casimir force is considered for real metals at
low temperatures. With the temperature decrease the mean free path for
electrons becomes larger than the field penetration depth. In this condition
description of metals with the impedance of anomalous skin effect is shown
to be more appropriate than with the permittivity. The effect is crucial for
the temperature correction. It is demonstrated that in the zero frequency
limit the reflection coefficients should coincide with those of ideal metal
if we demand the entropy to be zero at $T=0$. All the other prescriptions
discussed in the literature for the $n=0$ term in the Lifshitz formula give
negative entropy. It is shown that the temperature correction in the region
of anomalous skin effect is not suppressed as it happens in the plasma
model. This correction will be important in the future cryogenic
measurements of the Casimir force.
\end{abstract}

\section{\label{1}Introduction}

Attraction between parallel metallic plates predicted by Casimir in 1948 
\cite{Cas} (see \cite{BMM} for a recent review) has been measured in recent
experiments \cite{Lam1,AFM,HCM,Ederth,Chan,Onofrio} with high precision. To
calculate the force with the same precision, one has to take into account
different corrections to the original Casimir force \cite{BMM}. The
corrections due to finite conductivity of the plates and roughness of their
surfaces can be large but there is no principal problems with them and the
discussion is going around the material properties \cite
{Lam2,KRMM,BS2,LR,SL1,KMM,SL2}. However, the correction connected with the
finite temperature raised debate in the literature \cite
{BS,BGKM,GLR,SL3,KM,Lam3,Sirn,SB,BGKM2,TL,SL1,SL2}. The agreement between
different authors has not been reached yet.

In the early period the temperature correction was found for ideal metal. It
has been discovered that the correction following from the Lifshitz theory 
\cite{Lif} of electromagnetic fluctuations in nonhomogeneous media did not
agree with that found with different methods \cite{Meh,BM}. The problem
originates from the $n=0$ term in the Lifshitz formula giving the classical
contribution in the Casimir force from long wavelength fluctuations. Special
prescription for this term was proposed by Schwinger, DeRaad, and Milton 
\cite{SRM} to reconcile different approaches. Namely, one has to take the
limit of infinite permittivity before allowing the frequency go to zero.

The problem arose again when researches tried to find this correction for
real metals. It has been realized that the plasma and the Drude models for
the metal dielectric function gave different corrections, which did not
coincide when the Drude relaxation frequency $\omega _\tau $ is going to
zero. Direct application of the Lifshitz formula with the Drude dielectric
function gives large linear in temperature correction \cite{BS}, which
contradicts to the torsion pendulum experiment \cite{Lam1}. Negligible for
the experiments correction was found using the plasma model \cite{BGKM,GLR}.
However, the Drude model works much better for real metals and there is no a
satisfactory reason to justify the use of the plasma model.

It became obvious that something wrong with the $n=0$ term in the Lifshitz
formula and while a solid theoretical reason for modification of this term
is not found one has to use a reasonable prescription to define the $n=0$
term. It was proposed \cite{SL1} to apply for real metals the same
prescription as for ideal metal \cite{SRM}. It was motivated by the fact
that in the static limit the boundary conditions on the metal surface did
not depend on any material property. Later it has been shown \cite{SL3} that
the classical term can be constructed from very general principles using the
dimensional analysis, low frequency limit for the permittivity, and the
known result for ideal metal. The same conclusion was made in a recent paper 
\cite{TL}, where the authors stressed difference between boundary conditions
on dielectrics and metals in the zero frequency limit. In this case the
linear in temperature correction survives but it is suppressed by small
additional factor. The correction is unobservable \cite{SL1} in the
conditions of torsion pendulum experiment \cite{Lam1} but 3 times larger 
\cite{SL2} than the experimental errors in the atomic force microscope (AFM)
experiment \cite{HCM}. Note that it makes agreement between theory and
experiment better.

Quite different prescription was proposed for real metals by Klimchitskaya
and Mostepanenko \cite{KM}. The driving idea was to modify the reflection
coefficients in the $n=0$ term in such a way to make them coincide with that
of plasma model in the limit $\omega _\tau \rightarrow 0$. The reason for
that was based on the property of the reflection coefficient for
perpendicular polarization $r_1(\zeta ,q)$ considered as a function of two
variables: continuous imaginary frequency $\zeta $ and the absolute value of
the momentum along the plate $q$. It has integrated discontinuity in the
point $\zeta =0,\ q=0$ which authors \cite{KM} consider as unacceptable. The
resulting temperature dependence does not include the linear in temperature
term and negligible in conditions of the torsion pendulum and AFM
experiments.

Thus, at the moment there are three different approaches\ to the problem of
temperature dependence of the Casimir force. Recently an interesting
proposition \cite{BKM} to check the Nernst heat theorem for different
prescriptions discussed in the literature was\ made. According to the third
law of thermodynamics the entropy $S$ of the closed system in equilibrium
must go to zero in the limit $T\rightarrow 0$. Based on the analytical
result for the plasma model and numerical calculations for the Drude model a
conclusion was made that the prescription of Schwinger, Deraad, and Milton
cannot be applied to real metals because the entropy still finite in the $%
T\rightarrow 0$ limit. On the contrary, it was concluded that for the
prescription proposed by Klimchitskaya and Mostepanenko $S\rightarrow 0$
when $T\rightarrow 0$.

In this paper we intended to demonstrate that the real situation in the low
temperature limit is different. The reason for this is that with the
temperature decrease the mean free path for electrons $l$ increases and at
some temperature it inevitably becomes larger than the field penetration
depth $\delta $. It is the range of anomalous skin effect when the space
dispersion becomes important. Taking this effect into account modifies the
temperature correction, which does not coincide with that found in Ref.\cite
{BKM}. Independent interest to the correction at low temperatures is
stimulated by the progress in experiments which are going to cryogenic
temperatures to improve the precision \cite{HCM}.

The paper is organized as follows. In Section \ref{2} we give definition of
the Casimir free energy and explain the problem with the $n=0$ term. This
term is written out for all prescriptions discussed in the literature. In
Section \ref{3} it is demonstrated that neither plasma nor Drude models can
be used for real metals at low temperatures. The way to calculate the
temperature correction using the impedance of anomalous skin effect is
described. In Section \ref{4} the actual calculations are presented. The
results are applied to analyse the entropy behavior at $T\rightarrow 0$ and
to find the value of temperature correction in cryogenic range in Section 
\ref{5}. Our conclusions are collected in the last section.

\section{\label{2}Casimir free energy}

We start from the Lifshitz expression for the force per unit area between
two parallel plates separated by the distance $a$ at temperature $T$ \cite
{LP9}. Simple transformation of this formula allows to write the free energy
in the form \cite{BMM}

\begin{equation}
\label{Fenerg}{\cal F}(a,T)=\frac{kT}{8\pi a^2}{\sum\limits_{n=0}^\infty {}}%
^{\prime }\int\limits_{\xi _n}^\infty dy\,y\left[ \ln (1-r_1^2\left( \xi
_n,y\right) e^{-y})+r_1\rightarrow r_2\right] , 
\end{equation}

\noindent where the prime over the summation sign indicates that the first $%
n=0$ term has to be taken with the factor $1/2$ and

\begin{equation}
\label{taudef}\xi _n=n\tau ,\quad \tau =\frac{2\pi T}{T_{eff}},\quad
kT_{eff}=\frac{\hbar c}{2a}\equiv \hbar \omega _a. 
\end{equation}

\noindent Here $\xi _n$ are the dimensionless Matsubara frequencies $\xi
_n=\zeta _n/\omega _a$, where $\zeta _n=2\pi kTn/\hbar $. In Eq.(\ref{Fenerg}%
) $r_{1,2}^2\left( \zeta _n,q\right) $ are the reflection coefficients for
orthogonal ($r_1$) or parallel ($r_2$) polarizations. These coefficients
depend on the material via the dielectric function $\varepsilon \left(
i\zeta _n\right) $ at imaginary frequencies. In the Drude model $\varepsilon
\left( i\zeta _n\right) $ has the form

\begin{equation}
\label{Drude}\varepsilon \left( i\zeta \right) =1+\frac{\omega _p^2}{\zeta
\left( \zeta +\omega _\tau \right) } 
\end{equation}

\noindent and is defined by two parameters which are the plasma frequency $%
\omega _p$ and the relaxation frequency $\omega _\tau $. This model nicely
fits good metals in the whole frequency range excluding the interband
absorption region $\omega \sim \omega _p$, which does not give considerable
contribution especially at low temperatures.

Consider first the problematic $n=0$ term. Expression (\ref{Fenerg}) was
found by solving the equation for the Green function in nonhomogeneous media
at finite temperature \cite{LP9}. The explicit form of the reflection
coefficients $r_{1,2}^2$ depends on the boundary conditions for the Green
function at the metal-vacuum interface. Note that these conditions one can
put separately for any Matsubara components. Typically the continuity of
tangential components of electric and magnetic fields at the interface is
demanded. The Drude model gives for the amplitudes $r_{1,2}$

\begin{equation}
\label{Dref}r_1\left( \xi ,y\right) =\frac{\sqrt{R^2+y^2}-y}{\sqrt{R^2+y^2}+y%
},\quad r_2\left( \xi ,y\right) =\frac{\sqrt{R^2+y^2}-\left( 1+R^2/\xi
^2\right) y}{\sqrt{R^2+y^2}+\left( 1+R^2/\xi ^2\right) y}, 
\end{equation}

\noindent where the function $R(\xi )$ is defined as

\begin{equation}
\label{Rdef}R\left( \xi \right) =\frac{\omega _p}{\omega _a}\sqrt{\frac \xi {%
\xi +\omega _\tau /\omega _a}}. 
\end{equation}

\noindent In the $n=0$ term we have to put $\xi _0=0$ and the reflection
coefficients are

\begin{equation}
\label{presc1}r_1^2\left( 0,y\right) =0,\qquad r_2^2\left( 0,y\right) =1. 
\end{equation}

\noindent while for ideal metal both of these coefficients are equal to unit 
$r_1^2=r_2^2=1$. There is no way to reconcile the value of $r_1$ with that
of ideal metal since it does not depend on any parameters. This is why one
needs to introduce a prescription for the $n=0$ term. The coefficients (\ref
{presc1}) were used to calculate the temperature correction to the Casimir
force in Ref.\cite{BS}.

It was proposed \cite{SL1} to use the extension of the Schwinger, DeRaad,
and Milton prescription to real metals. Since in the static field limit the
boundary condition on the metal surface does not depend on the
characteristics of a particular metal, the reflection coefficients were
changed with that of ideal metal

\begin{equation}
\label{presc2}r_1^2\left( 0,y\right) \rightarrow 1,\qquad r_2^2\left(
0,y\right) =1. 
\end{equation}

\noindent Different prescription was introduced by Klimchitskaya and
Mostepanenko \cite{KM} who proposed to modify $r_1^2$ in such a way that in
the limit $\omega _\tau \rightarrow 0$ it coincided with the plasma model
result

\begin{equation}
\label{presc3}r_1^2\left( 0,y\right) \rightarrow r_1^2\left( y,y\right)
=\left( \frac{\sqrt{R^2\left( y\right) +y^2}-y}{\sqrt{R^2\left( y\right) +y^2%
}+y}\right) ^2,\qquad r_2^2\left( 0,q\right) =1. 
\end{equation}

\noindent The motivation of (\ref{presc2}) and (\ref{presc3}) was briefly
described above and we are not going to the details referring to the
original papers \cite{SL3,KM}.

Let us denote the $n=0$ term in (\ref{Fenerg}) as ${\cal F}_0\left(
a,T\right) $. Calculating the corresponding integral with the reflection
coefficients (\ref{presc1}-\ref{presc3}) one gets

\begin{equation}
\label{zterm}{\cal F}_0\left( a,T\right) =-\alpha \frac{kT}{8\pi a^2}\zeta
\left( 3\right) , 
\end{equation}

\noindent where $\zeta \left( x\right) $ is the zeta function. The
coefficient $\alpha $ has the following values for the prescriptions (\ref
{presc1}), (\ref{presc2}), and (\ref{presc3}), respectively:

$$
\alpha ^{\left( i\right) }=\frac 12,\qquad \alpha ^{\left( ii\right) }=1, 
$$

\begin{equation}
\label{alpha}\alpha ^{\left( iii\right) }=1-4\frac{\omega _a}{\omega _p}-%
\frac{\omega _\tau }{\omega _p}\frac 2{\zeta \left( 3\right) }I_2\left( 
\frac{\omega _\tau }{\omega _a}\right) +O\left( \frac{\omega _a^2}{\omega
_p^2}\right) . 
\end{equation}

\noindent The function $I_2\left( x\right) $ varies slowly with $x$ and is
given by Eq.(68) \cite{KM}.

\section{\label{3}Anomalous skin effect}

Let us consider now the $n\geq 1$ terms in (\ref{Fenerg}) when the
temperature is going down. The plasma frequency $\omega _p$ is defined by
the electron density in metal and practically does not depend on
temperature. On the contrary, the relaxation frequency $\omega _\tau $
changes with $T$ significantly. If $T<T_D$, where $T_D$ is the Debye
temperature for a given metal, then the dependence $\omega _\tau \left(
T\right) $ can be presented as (see, for example, \cite{Abr})

\begin{equation}
\label{OmT}\omega _\tau \left( T\right) =\omega _\tau \left( 0\right)
+C_eT^2+C_{ph}T^5 
\end{equation}

\noindent with the parameters $\omega _\tau (0),\ C_e,$ and $C_{ph}$. Here
the first term is connected with the scattering on lattice irregularities
and impurities, the second one describes scattering on electrons, and the
third term corresponds to the scattering on phonons. The first and second
terms can dominate only at very low temperatures. The first term needs
special discussion. The relaxation frequency is proportional to the material
resistivity which disappears at $T\rightarrow 0$ for perfect monocrystals.
If we are going to check the third law of thermodynamics for our system, we
have to take this equilibrium state and choose $\omega _\tau (0)=0$. To
describe the temperature behavior of real material used in the experiment,
it can be wrong. For evaporated metallic films the residual resistivity can
be significant due to large density of defects but it is much smaller than
the resistivity at room temperature.

One can easily see the result of Ref.\cite{BKM} without any calculations. At
low temperature $\omega _\tau /\omega _a$ in (\ref{Rdef}) is going to zero
faster than $\xi _n\sim T$. Therefore, for all $n\neq 0$ terms one can use
the plasma model. The reflection coefficients in the $n=0$ term (\ref{presc3}%
) were prescribed to reproduce the plasma model at $\omega _\tau \rightarrow
0$. Therefore, the plasma model completely describes the situation at low
temperatures. It is well known \cite{BGKM} that the leading temperature
correction in this model behaves as $T^3$ and the entropy will go to zero as 
$S\sim T^2$. Any other prescription will not agree with the Nernst theorem
since the $n=0$ term does not coincide with that for the plasma model. For (%
\ref{presc1}) the entropy becomes negative and for (\ref{presc2}) it is
positive but both of them are finite at $T=0$. No doubt that at low
temperature and for equilibrium state the relaxation frequency becomes
negligible. However, it is wrong to think that the reflection coefficients (%
\ref{Dref}) for $n\neq 0$ will be the same at low temperatures. This is
because the mean free path for electrons increases with the temperature
decrease according to the relation $l=v_F/\omega _\tau (T)$, where $v_F$ is
the Fermi velocity. At some small temperature inevitably the relation $l\gg
\delta $ will be fulfilled, where $\delta =c/\omega _p$ is the field
penetration depth\footnote{%
Strictly speaking $\delta $ depends on frequency. We assumed that $\omega
=\zeta _n\gg \omega _\tau \left( T\right) $ ($n\neq 0$). If it is not the
case, we have to take $\delta =(c/\omega _p)\sqrt{2\omega _\tau /\omega }$
but the qualitative conclusion will not change.}. Then the condition $l\gg
\delta $ is equivalent to the following

\begin{equation}
\label{ASE}\omega _\tau \left( T\right) \ll \frac{v_F}c\omega _p\equiv
\Omega . 
\end{equation}

\noindent The frequency $\Omega =(v_F/c)\omega _p$ is often used as a
characteristic frequency of anomalous skin effect.

Thus, the local connection between the current and electric field, which is
true for $l<\delta $, is broken at low temperature and space dispersion
becomes important. This is the range of anomalous skin effect \cite{LP10},
when the dielectric function $\varepsilon (\omega )$ cannot be used any more
for description of the metal. Instead the interaction with the field is
defined by the surface impedance $Z(\omega )$ \cite{LL8,LP10}. The relation $%
Z(\omega )\sim \varepsilon ^{-1/2}(\omega )$, which holds for the normal
skin effect, is broken.

When the metal is described by the impedance the boundary condition on its
surface will be \cite{LL8}

\begin{equation}
\label{Leont}{\bf E}_t=Z\left( \omega \right) \left( {\bf H}_t\times {\bf n}%
\right) , 
\end{equation}

\noindent where ${\bf E}_t{\bf ,H}_t$ are the tangential components of
electric and magnetic fields, ${\bf n}$ is the unit vector normal to the
surface and directed inside of the metal. It holds true while the impedance
is small. If this boundary condition is used for the Matsubara components ($%
n\geq 1$) of the Green function, then the reflection coefficients can be
written as follows \cite{MT,BKR}

\begin{equation}
\label{reflim}r_1=\frac{\xi _n-yZ\left( i\zeta _n\right) }{\xi _n+yZ\left(
i\zeta _n\right) },\quad r_2=\frac{y-\xi _nZ\left( i\zeta _n\right) }{y+\xi
_nZ\left( i\zeta _n\right) }. 
\end{equation}

\noindent Impedance in the range of anomalous skin effect one can find
solving the kinetic equation for not in equilibrium distribution function of
electrons in the electromagnetic wave \cite{LP10} or using simple
qualitative analysis \cite{Abr}.

In the frequency range $\omega <\Omega $ (strong anomalous skin effect) the
impedance is \cite{LP10,Abr}

\begin{equation}
\label{Zstr}Z\left( \omega \right) =e^{-i\pi /3}\left( \frac vc\frac{\omega
^2}{\omega _p^2}\right) ^{1/3}. 
\end{equation}

\noindent Here $v=\beta v_F$, where $\beta $ is a factor $\sim 1$, which is
defined by the structure of the Fermi surface. Note that the relaxation
frequency falls out from the impedance at all. This is because the field
interacts mainly with electrons, which are moving in the surface layer of
thickness $\delta $. For this reason significant is the effective
conductivity $\sigma _{eff}=\sigma \delta /l$, which does not depend on $%
\omega _\tau $. Actually we are interesting in the impedance at imaginary
frequencies. For the strong anomalous skin effect it will be

\begin{equation}
\label{Zim}Z\left( i\zeta _n\right) =\left( \frac vc\frac{\omega _a^2}{%
\omega _p^2}\xi _n^2\right) ^{1/3}. 
\end{equation}

\noindent This expression is true for $\xi _n<\Omega /\omega _a$ while the
range of important $n$ in the sum (\ref{Fenerg}) is given by the condition $%
\xi _n\sim 1$. For small $\Omega /\omega _a$ the relation (\ref{Zim}) cannot
be used for all important $n$. However, as we will see below the temperature
dependent part of free energy always can be described by (\ref{Zim}).

The anomalous skin effect will change not only the temperature correction
but the main temperature independent part of the Casimir force or free
energy. To find this change one has to know the impedance in the whole range
of $\xi _n$. It is very important for comparison with the experimental data
in cryogenic range but out of the scope of this paper and will be discussed
elsewhere.

\section{\label{4}Calculation of the free energy}

Now we are ready to find the sum of all $n\geq 1$ terms in (\ref{Fenerg}).
This sum can be transformed using the Abel-Plana formula \cite{MTRev}. To
avoid inconvenient integrals we first change the summation index to $m=n-1$.
In this way the problematic coefficient $r_{1}(0,y)$ will not appear in the
formula as it usually happens \cite{BGKM}. The resulting expression is
longer than usual but simpler for calculations. Separating the temperature
independent part one finds

\begin{equation}
\label{Fng1}{\cal F}(a,T)-{\cal F}_0(a,T)={\cal F}(a,0)+\frac{kT}{8\pi a^2}%
\left[ \frac 12\left( I_1^{(1)}+I_1^{(2)}\right) -\left(
I_2^{(1)}+I_2^{(2)}\right) +\left( I_3^{(1)}+I_3^{(2)}\right) \right] . 
\end{equation}

\noindent Here the temperature independent term ${\cal F}(a,0)$ is

\begin{equation}
\label{FT0}{\cal F}(a,0)=\frac{\hbar c}{16\pi a^3}\int\limits_0^\infty d\xi
\int\limits_0^\infty dy\left[ \left( \xi +y\right) \ln \left( 1-r_1^2\left(
\xi ,y+\xi \right) e^{-y-\xi }\right) +r_1\rightarrow r_2\right] 
\end{equation}

\noindent and the integrals $I_k^{(j)}\ (j=1,2;\ k=1,2,3)$ are defined as

$$
I_1^{(j)}=\int\limits_\tau ^\infty dyy\ln \left[ 1-r_j^2\left( \tau
,y\right) e^{-y}\right] , 
$$

$$
I_2^{(j)}=\int\limits_0^1dt\int\limits_{\tau t}^\infty dyy\ln \left[
1-r_j^2\left( \tau t,y\right) e^{-y}\right] , 
$$

\begin{equation}
\label{intsdef}I_3^{(j)}=2Im\int\limits_0^\infty \frac{dt}{e^{2\pi t}-1}%
\int\limits_\tau ^\infty dy\left( y+i\tau t\right) \ln \left[ 1-r_j^2\left(
\tau +i\tau t,y+i\tau t\right) e^{-y-i\tau t}\right] . 
\end{equation}

\noindent In the temperature independent term (\ref{FT0}) the important
range of continuous variable $\xi $ is $\xi \sim 1$ and, as was mentioned
above, the impedance (\ref{Zim}) cannot be used to cover all the range of $%
\xi $. But this term is not interesting for us. In the temperature dependent
terms $\xi =\tau t\sim \tau $ as one can see from (\ref{intsdef}). We can
use (\ref{Zim}) in the integrals (\ref{intsdef}) if $\tau <\Omega /\omega _a$
or equivalently $2\pi kT<\hbar \Omega $. This relation together with $T<T_D$
will be supposed to be true.

In all the integrals (\ref{intsdef}) the lower limit in $y$ can be changed
into zero. Really, the error connected with such a replacement will be of
the order of $\tau ^2\ln \tau $. It will be considered as small for
analytical calculations but will be taken into account in the numerical
procedure. Then the integrals $I_k^{(1)}$ depending on the reflection
coefficient $r_1^2$ (\ref{reflim}) will be defined only by the parameter

\begin{equation}
\label{A1}A=\left( \frac cv\frac{\omega _p^2}{\omega _a^2}\tau \right)
^{1/3}. 
\end{equation}

\noindent Even for small $\tau $ the typical value of this parameter is
large $A\gg 1$. Only for very low temperature $kT\ll (\omega _a/\omega
_p)^3\hbar \Omega /2\pi $ it becomes small ($A\ll 1$). It is in contrast
with the parameter $B$ which defines the integrals $I_k^{(2)}$

\begin{equation}
\label{A2}B=\left( \frac vc\frac{\omega _a^2}{\omega _p^2}\tau ^5\right)
^{1/3}. 
\end{equation}

\noindent This parameter is always small in the range of interest $\tau \ll
1 $.

\subsection{Free energy in the limit $A\ll 1$}

Let us analyse now the analytical dependence of the integrals (\ref{intsdef}%
) from the parameters $A,\,B$ at very low temperatures when $A\ll 1$. We
start from $I_1^{(1)}$. Neglecting the correction $\sim \tau ^2\ln \tau $
this integral can be written

\begin{equation}
\label{I1repr}I_1^{(1)}=\int\limits_0^\infty dyy\ln \left( 1-e^{-y}\right)
+\int\limits_0^\infty dyy\ln \left[ 1+\frac{4Ay}{\left( y+A\right) ^2}\left(
e^y-1\right) ^{-1}\right] . 
\end{equation}

\noindent The first integral here is equal $-\zeta \left( 3\right) $. It
describes contribution of the ideal metal. For the second integral one can
find the leading terms in the limit $A\rightarrow 0$. The important
contribution gives the range $y\ 
\raisebox{-0.5ex}{$\stackrel {\scriptstyle <}{\scriptstyle \sim}$}\ 1$ but
the main one is collected near $y\sim \sqrt{A}$. Then $I_1^{(1)}$ can be
presented as

$$
I_1^{(1)}\approx -\zeta \left( 3\right) +\int\limits_0^1dyy\ln \left( 1+%
\frac{4A_1}{y^2}\right) . 
$$

\noindent The result is the following

\begin{equation}
\label{I1res}I_1^{(1)}=-\zeta \left( 3\right) -2A\left( \ln A+2\ln
2-1\right) +O\left( A^2\right) . 
\end{equation}

\noindent The integral $I_2^{(1)}$ can be estimated in the same way but one
has to change $A\rightarrow A\,t^{1/3}$ and then integrate it over $t$. It
gives

\begin{equation}
\label{I2res}I_2^{(1)}=-\zeta \left( 3\right) -\frac 32A\left( \ln A+2\ln 2-%
\frac 54\right) +O\left( A^2\right) . 
\end{equation}

The same procedure can be applied for $I_3^{(1)}$ but realization is more
complicated. Contribution of the ideal metal (zero impedance) is negligible
in our approximation since the leading term is known \cite{Meh,BM} to be $%
\tau ^2$. Correction due to nonzero impedance depends on $A$. In the same
approximation as for (\ref{I1res}) and (\ref{I2res}) one has

$$
I_3^{(1)}=2Im\int\limits_0^\infty \frac{dt}{e^{2\pi t}-1}\int\limits_0^1dy%
\left( y+i\tau t\right) \ln \left( 1+\frac{4A\,\left( 1+it\right) ^{1/3}}{y^2%
}\right) . 
$$

\noindent The inner integral is calculated analytically. The contribution
from the term proportional to $i\tau t$ will be of the order of $\tau \sqrt{A%
}$ which is small in comparison with the leading terms. All the rest can be
written as

\begin{equation}
\label{I3res}I_3^{(1)}=4A(q_1\ln A+q_2)+O\left( \tau \sqrt{A}\right) . 
\end{equation}

\noindent The constants $q_{1,2}$ here are

\begin{equation}
\label{qdef}q_1=\int\limits_0^\infty \frac{dt\left( 1+t^2\right) ^{1/6}\sin
\vartheta }{e^{2\pi t}-1},\quad q_2=\int\limits_0^\infty \frac{dt\left(
1+t^2\right) ^{1/6}}{e^{2\pi t}-1}\left[ \sin \vartheta \left( \ln 4\left(
1+t^2\right) ^{1/6}-1\right) +\vartheta \cos \vartheta \right] , 
\end{equation}

\noindent where $\tan 3\vartheta =t$. Numerically these constants are $%
q_1=0.0137$, $q_2=0.0191$.

The integrals $I_k^{(2)}$ can be calculated quite similar. Since $B$ is
always much smaller than $A$ one can completely neglect the terms containing 
$B$ and for $I_k^{(2)}$ one has

\begin{equation}
\label{I(2)def}I_1^{(2)}=-\zeta \left( 3\right) ,\quad I_2^{(2)}=-\zeta
\left( 3\right) ,\quad I_3^{(2)}=0. 
\end{equation}

\noindent Substituting (\ref{I1res}-\ref{I3res}, \ref{I(2)def}) into (\ref
{Fng1}) and taking into account explicit expression (\ref{zterm}) for ${\cal %
F}_0(a,T)$ one finds the temperature correction to the free energy in the
limit $A\rightarrow 0$

\begin{equation}
\label{FsmallA}\Delta {\cal F}\left( a,T\right) ={\cal F}\left( a,T\right) -%
{\cal F}\left( a,0\right) =\frac{kT}{8\pi a^2}\left[ \left( 1-\alpha \right)
\zeta \left( 3\right) +A\left( \left( \frac 12+4q_1\right) \ln A+\ln 2-\frac 
78+4q_2\right) \right] . 
\end{equation}

\noindent This expression is true for very low temperature (see Eq.(\ref{A1}%
)) and is appropriate only to check the Nerst theorem. It will be discussed
in the next section. For realistic low temperatures, which will be explored
in the near future experiments, the typical value of $A$ is large even for
small $\tau $ and we have to analyse the opposite limit $A\gg 1$.

\subsection{Free energy in the limit $A\gg 1$}

To calculate the integrals $I_k^{(1)}$ in the temperature range when the
parameter $A$ is large but the temperature still small ($\tau \ll 1$), it
will be convenient again to separate the contribution of ideal metal and
neglect the correction $\sim \tau ^2\ln \tau $ due to change of the low
limit in integrals into zero. Then for $I_1^{(1)}$ one has the same
representation (\ref{I1repr}) and similar for the other integrals. Because $%
A $ is large one can expand the logarithm in series and collect the
coefficient at different powers of $1/A$. For $I_1^{(1)}$ this procedure can
be done straightforward and we find for the first two terms

\begin{equation}
\label{I1(1)large}I_1^{(1)}=-\zeta \left( 3\right) +8\zeta \left( 3\right)
\left( \frac 1A-\frac 6{A^2}\right) . 
\end{equation}

In case of $I_2^{(1)}$ the role of $A$ plays $A\,t^{1/3}$. In this integral
after logarithm expansion one has to make first the integration over $t$ not
to run onto divergencies. The result is the following

\begin{equation}
\label{I2(1)large}I_2^{(1)}=-\zeta \left( 3\right) +12\zeta \left( 3\right)
\left( \frac 1A-\frac{12}{A^2}\right) . 
\end{equation}

\noindent In the third integral instead of $A$ appears $A\,(1+it)^{1/3}$. In
our approximation one can completely neglect $i\tau t$ in comparison with $A$
or with $y$. Making the same procedures one gets

\begin{equation}
\label{I3(1)large}I_3^{(1)}=16\zeta \left( 3\right) \left( \frac{p_1}A-\frac{%
6p_2}{A^2}\right) , 
\end{equation}

\noindent where the coefficients are defined as

\begin{equation}
\label{pdef}p_1=\int\limits_0^\infty dt\frac{\sin \vartheta }{\left( e^{2\pi
t}-1\right) \left( 1+t^2\right) ^{1/6}},\quad p_2=\int\limits_0^\infty dt%
\frac{\sin 2\vartheta }{\left( e^{2\pi t}-1\right) \left( 1+t^2\right) ^{1/3}%
}. 
\end{equation}

\noindent Here the angle $\vartheta $ is defined as in (\ref{qdef}).
Numerically the coefficients are $p_1=0.0133,\ p_2=0.0262$.

The integrals $I_k^{(2)}$giving the contribution of parallel ($r_2$)
polarization depends on the parameter $B$ which is small while $\tau \ll 1$.
Therefore, they give the same result (\ref{I(2)def}) as in the case of small 
$A$. Collecting all together one finds the temperature correction to the
free energy in the limit $A\gg 1$

\begin{equation}
\label{FlargeA}\Delta {\cal F}\left( a,T\right) =\frac{kT}{8\pi a^2}\zeta
\left( 3\right) \left[ \left( 1-\alpha \right) -8\left( \frac{1-2p_1}A-\frac{%
15-12p_2}{A^2}\right) \right] . 
\end{equation}

\noindent The temperature correction to the Casimir force can be easily
found via $\Delta {\cal F}$. Correction to the force between sphere and
plate $\Delta F_{sp}$ is directly proportional to $\Delta {\cal F}$ if we
adopt the proximity force theorem \cite{Dzjal}

\begin{equation}
\label{Fsp}\Delta F_{sp}\left( a,T\right) =2\pi R\,\Delta {\cal F}\left(
a,T\right) , 
\end{equation}

\noindent where $R$ is the sphere radius. The temperature correction for the
force between two plates $\Delta F_{pp}$ is

\begin{equation}
\label{Fpp}\Delta F_{pp}\left( a,T\right) =-\frac \partial {\partial a}%
\Delta {\cal F}\left( a,T\right) =\frac{kT}{4\pi a^3}\zeta \left( 3\right)
\left[ \left( 1-\alpha \right) -8\left( \frac 32\frac{1-2p_1}A-2\frac{%
15-12p_2}{A^2}\right) \right] 
\end{equation}

Thus, we found the asymptotics for $\Delta {\cal F}\left( a,T\right) $ in
the limits $A\rightarrow 0$ (\ref{FsmallA}) and $A\rightarrow \infty $ (\ref
{FlargeA}). Now let us calculate numerically this function in the transition
region.

\subsection{Numerical calculation of the free energy}

For numerical calculation we parametrize the temperature correction to the
free energy in the following way

\begin{equation}
\label{Fparam}\Delta {\cal F}\left( a,T\right) =\frac{kT}{8\pi a^2}\left[
\left( 1-\alpha \right) \zeta \left( 3\right) -G\left( A,\tau \right)
\right] . 
\end{equation}

\noindent The absolute scale of the correction is given by the factor $%
kT/8\pi a^2$ then $G(A,\tau )$ can be treated as the relative temperature
correction if $\alpha =1$. The function $G(A,\tau )$ is calculated
numerically as a function of $A$ for a few given values of $\tau $. There is
no need to separate $B$ as an independent parameter since it is connected
with the other two as $B=\tau ^2/A$. Alternatively at fixed material
parameters $\omega _p$ and $v$ it can be found as a function of $T$ at a
fixed $a$. Actual calculation has been done using the relation

$$
G\left( A,\tau \right) =-\frac 12\left( I_1^{(1)}+I_1^{(2)}-2\zeta \left(
3\right) \right) +\left( I_2^{(1)}+I_2^{(2)}-2\zeta \left( 3\right) \right)
-\left( I_3^{(1)}+I_3^{(2)}\right) . 
$$

\begin{figure}
\centerline{\epsffile{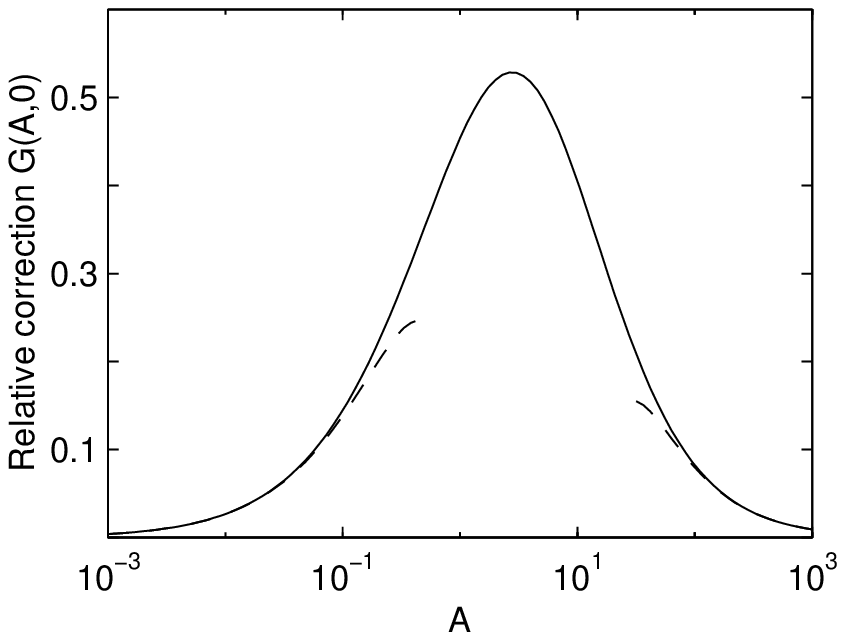}}
\caption{The relative temperature correction $G(A,0)$ as a function of the 
parameter $A$ (\protect \ref{A1}). The other parameter $\tau $ 
(\protect \ref{taudef}) is assumed to be small. The solid curve represents 
numerical calculation. The dashed lines correspond to the asymptotics 
(\protect \ref{FsmallA}) and (\protect \ref{FlargeA}).}

\label{fig1} 

\end{figure}

\begin{figure}
\centerline{\epsffile{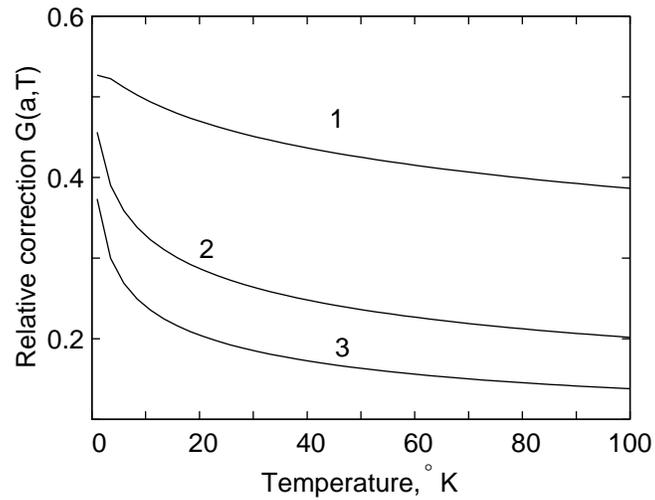}}
\caption{The relative temperature correction $G(a,T)$ as a function of temperature
at a few values of separations between bodies: (1) $a=100\ nm$, (2) $a=300\ nm$,
(3) $a=500\ nm$.} 

\label{fig2} 

\end{figure}

\noindent The integrals were calculated according to (\ref{intsdef}) with
the absolute precision of $10^{-6}$. Fig.\ref{fig1} shows the function $%
G\left( A,0\right) $ and its asymptotics at $A\rightarrow 0$ and $%
A\rightarrow \infty $, which can be extracted from (\ref{FsmallA}) and (\ref
{FlargeA}). One can see that there is good agreement of numerical
calculation and analytical asymptotics. If we consider nonzero but small $%
\tau $, the result will change only slightly due to correction $\sim \tau
^2\ln \tau $. For $\omega _p=1.37\cdot 10^{16}\ rad/s$ and $v=1.5\cdot 10^8\
cm/s$ (gold) $G(a,T)$ as a function of $T$ is shown in Fig.\ref{fig2} at a
few values of $a$. This figure shows that the temperature correction is
always significant in the interesting range of temperatures and distances
between plates.

\section{\label{5}Discussion of the results}

\subsection{Entropy in the limit $T\rightarrow 0$}

We already saw that the anomalous skin effect changes behavior of the free
energy at low temperatures. Now we are able to answer the question which of
the prescriptions (\ref{presc1}), (\ref{presc2}) or (\ref{presc3}) agrees
with the third law of thermodynamics. In the low temperature limit entropy
can be calculated from (\ref{FsmallA})

\begin{equation}
\label{Entropy}S=-\frac{\partial {\cal F}}{\partial T}=\frac k{8\pi a^2}%
\left[ \left( \alpha -1\right) \zeta \left( 3\right) -\frac 43A\left( \left( 
\frac 12+4q_1\right) \ln A+\ln 2-\frac 34+4q_2+q_1\right) \right] . 
\end{equation}

\noindent Parameter $\alpha $ corresponding to different prescriptions used
for the $n=0$ term in the Lifshitz formula is given by Eq.(\ref{alpha}). If
this term is calculated as it appears in the Lifshitz formula \cite{BS}
without any modification, then $\alpha =\alpha ^{\left( i\right) }=1/2$. In
this case the entropy is finite at $T=0$ ($A=0$) and, moreover, it is
negative. This conclusion coincides with that made in Ref.\cite{BKM}, where
at low temperatures the authors described the plate material with the plasma
model. It can be considered as an additional physical argument that
unmodified Lifshitz formula cannot be used to get temperature behavior of
the Casimir force. The previous argument \cite{SL3,KM} was that in this
approach the $n=0$ term in (\ref{Fenerg}) does not depend on any material
parameter (see (\ref{zterm}) with $\alpha =\alpha ^{\left( i\right) }$ ) and
for this reason it cannot be reconciled with the ideal metal result, which
is two times larger. These arguments show that the Lifshitz formula is
really in trouble. It obviously has to be modified but after a few years of
active discussion still there is no solid theoretical understanding what is
wrong and how one can do this.

Prescription proposed by Klimchitskaya and Mostepanenko \cite{KM} was
designed to reproduce the plasma model result for the $n=0$ term in the
limit $\omega _\tau \rightarrow 0$. This limit was supposed to be realized
at low temperatures but, as was explained above, the material has to be
described in this case by the impedance of anomalous skin effect rather than
the plasma model permittivity. For this reason the entropy for $\alpha
=\alpha ^{\left( iii\right) }$ is not going to zero when $T\rightarrow 0$.
It is easy to see from (\ref{Entropy}) and (\ref{alpha}) that $S$ is finite
and negative at $T=0$. Therefore, this prescription is also does not obey
the Nernst theorem. At higher temperatures, when the Drude model is valid,
the $n=0$ term given by (\ref{zterm},\ref{alpha}) depends separately from $%
\omega _p$ and $\omega _\tau $. Physically it is unacceptable since at low
frequencies the only parameter characterizing a metal in respect to
electromagnetic field is the conductivity $\sigma \sim \omega _p^2/\omega
_\tau $ \cite{LL8}.

If we use the prescription (\ref{presc2}) \cite{SL1} then $\alpha =\alpha
^{\left( ii\right) }=1$. In this case the temperature independent term in (%
\ref{Entropy}) is canceled completely and entropy is going to zero with the
temperature as $T^{1/3}\ln T$. It is the only prescription which is in
agreement with the third law of thermodynamics. This prescription was
criticized \cite{BGKM,KM} on the basis that the $n=0$ term does not depend
on any material parameter (see (\ref{zterm})). We already explained \cite
{SL3} that in the static (long wavelength) limit the boundary condition for
metals do not include a particular metal parameters. More specifically, from
dimensional analysis it follows that $\alpha $ in (\ref{zterm}) can be a
function of the only variable \cite{SL3}

\begin{equation}
\label{alfun}\alpha =\alpha \left( \frac{\omega _\tau \omega _a}{\omega _p^2}%
\right) . 
\end{equation}

\noindent The characteristic frequency which appears here is huge $\omega
_p^2/\omega _\tau \sim 10^{18}\ rad/s$ (for good metals at room
temperature). Of course, one can expand $\alpha $ in a series

$$
\alpha \approx \alpha \left( 0\right) +\alpha ^{\prime }\left( 0\right)
\left( \frac{\omega _\tau \omega _a}{\omega _p^2}\right) 
$$

\noindent but the correction to $\alpha \left( 0\right) $ becomes important
only at microscopic distances between the plates $a\sim 10^{-8}\ cm$ where
the macroscopic Casimir force is not defined. The value of $\alpha (0)$
should coincide with that for ideal metal which is known to be 1. In this
way the prescription (\ref{presc2}) is reproduced without direct reference
to the Drude reflection coefficients (\ref{Dref}).

\subsection{Temperature correction to the Casimir force}

The results of previous section for the temperature correction to the free
energy or equivalently to the force (see (\ref{Fsp}) and (\ref{Fpp})) show
that for $\alpha =1$ the correction always increases the absolute value of
the force (the correction is negative so as the attractive force). The
correction is relatively large $\sim kT/8\pi a^2$. It is unlike to the
plasma model where additional small factors $\tau ^2$ \cite{BGKM} or $\omega
_a/\omega _p$ \cite{SL3} appear if one uses prescriptions (\ref{presc3}) or (%
\ref{presc2}), respectively. In this sense one can say that in the range of
anomalous skin effect the temperature correction becomes large. The relative
correction $G(a,T)$ in (\ref{Fparam}) is maximal ($\approx 0.53$) at the
temperature

\begin{equation}
\label{Tmax}kT_m\approx 18\frac{\hbar \omega _a}{2\pi }\left( \frac vc\frac{%
\omega _a^2}{\omega _p^2}\right) . 
\end{equation}

\noindent In the ideal metal limit $\omega _p\rightarrow \infty $ this
temperature is going to zero and instead of large correction appears usual
ideal metal correction \cite{Meh,BM} given by

\begin{equation}
\label{idmet}G\left( a,T\right) =\zeta \left( 3\right) \left( \frac \tau {%
2\pi }\right) ^2-\frac{\pi ^3}{45}\left( \frac \tau {2\pi }\right) ^3. 
\end{equation}

Let us describe the temperature range where the result of this work will be
applicable. Consider first the condition $l\gg \delta $ which guarantees
that anomalous skin effect plays important role. It can give us the upper
limit on the temperature. We approximate $\omega _\tau (T)$ with the
Bloch-Gr\"uneisen formula

\begin{equation}
\label{BG}\frac{\omega _\tau (T)}{\omega _\tau (T_0)}=\left( \frac T{T_0}%
\right) ^5\frac{F_5\left( T/T_D\right) }{F_5\left( T_0/T_D\right) },\quad
F_5\left( T/T_D\right) =\int\limits_0^{T/T_D}dx\frac{x^5}{\left(
e^x-1\right) \left( 1-e^{-x}\right) }, 
\end{equation}

\noindent where $T_0$ is some fixed temperature, for example, $T_0=0^{\circ
}\,C$. This formula does not take into account scattering on the defects and
electrons, which can be important at very low temperatures, but it is good
to find the upper limit on $T$. The value of $\omega _\tau (T_0)$ can be
fixed via the material resistivity $\rho $ using the relation $\omega _\tau
=\varepsilon _0\omega _p^2\rho $, where $\varepsilon _0$ is the permittivity
of vacuum. For gold parameters $\omega _p=1.37\cdot 10^{16}\ rad/s$, $%
v_F=1.4\cdot 10^8\ cm/s$, $\rho (T_0)=2.06\ \mu \Omega \cdot cm$ we found $%
l/\delta >5$ for $T<113^{\circ }\,K$ and $l/\delta >10$ for $T<67^{\circ
}\,K $. The other condition $2\pi kT<\hbar \Omega $ ensures that a specific
expression for the impedance (\ref{Zim}) will be true. It restricts the
temperature by the value $T<77.5^{\circ }\,K$ (gold). Therefore, already for
liquid nitrogen our result for the temperature correction is applicable.

\section{\label{6}Conclusion}

We considered the temperature correction to the Casimir free energy (force)
in the low temperature range. The aims of this analysis were to investigate
the behavior of entropy in the limit $T\rightarrow 0$ and find out if it is
possible to neglect the temperature correction in the future low temperature
experiments. The main observation of this work is that at low temperature $%
T<T_D$ (Debye temperature) neither plasma nor Drude models for material
permittivity are good for real metals. The reason is that with the
temperature decrease the mean free path for electrons becomes larger than
the penetration depth for electromagnetic field in metal. This relation
between parameters is realized for the anomalous skin effect for which
description of metals with impedance is more appropriate than with the
dielectric function. This change in the description is important for the
temperature correction.

It is known that the first $n=0$ term in the Lifshitz formula for the
Casimir force is controversial. A lot of discussion in the literature is
going on around this term. The problem is how one can define this term
correctly. At the moment there are three different approaches corresponding
to three different results for the correction. It was proposed to use the
third law of thermodynamics to choose one of the approach \cite{BKM}. We
analysed the entropy behavior in the limit $T\rightarrow 0$ for discussed in
the literature prescriptions and came to a conclusion that one has to define
the reflection coefficients in the static field limit as for the ideal metal 
$r_1^2\rightarrow 1,\ r_2^2=1$ to get agreement with the third law of
thermodynamics. It is in contrast with the conclusion of Ref. \cite{BKM},
where the plasma model was used for metals at low temperatures.

To reduce the noise, the future experiments on precise measurement of the
Casimir force will explore the low temperature range. Previously metals at
low temperatures were described by the plasma model because the Drude
relaxation frequency decreases fast with temperature. The plasma model
predicts a negligible temperature correction to the force. We demonstrated
that the anomalous skin effect makes drastic change. In all interesting from
the experimental point of view range of temperatures and distances between
bodies the correction is not negligible on the precision level of modern
experiments.

We do not consider our result as final solution of the problem with the
temperature correction. This is because a real solution should not be based
on any prescription. The problem indicates that something wrong with the
Lifshitz formula and efforts have to be directed on the careful analysis of
this formula. It becomes more and more clear that the essence of the problem
lies in the boundary conditions on the metal surface. The impedance
condition (\ref{Leont}) applied to the $n=0$ term reproduced the right
result for the reflection coefficients $r_1^2=r_2^2=1$ \cite{TL}. However,
it is not clear why there is no a smooth transition between continuity of
tangential components of ${\bf E}$ and$\ {\bf H}$ and the impedance boundary
condition.

\newpage

\end{document}